\documentclass[sigconf]{acmart}

\usepackage{graphicx}
\usepackage[utf8]{inputenc} 
\usepackage[T1]{fontenc}    
\usepackage{hyperref}       
\usepackage{url}            
\usepackage{booktabs}       
\usepackage{amsfonts}       
\usepackage{nicefrac}       
\usepackage{microtype}      
\usepackage{csvsimple}
\usepackage{booktabs} 
\usepackage{siunitx} 
\usepackage{makecell}
\AtBeginDocument{%
  \providecommand\BibTeX{{%
    \normalfont B\kern-0.5em{\scshape i\kern-0.25em b}\kern-0.8em\TeX}}}

\setcopyright{acmcopyright}
\copyrightyear{2018}
\acmYear{2018}
\acmDOI{XXXXXXX.XXXXXXX}

\acmConference[KDD '22]{28th ACM SIGKDD Conference On Knowledge Discovery and Data Mining}{Aug 14--18,2022}{Washington, DC, USA}
\acmPrice{}
\acmISBN{}

\begin{document}


\title{Too Fine or Too Coarse? The Goldilocks Composition of Data Complexity for Robust Left-Right Eye-Tracking Classifiers}

\author{Brian Xiang}
\authornote{Both authors contributed equally to this research.}
\email{bxiang1@swarthmore.edu}
\orcid{0000-0002-3403-2014}
\affiliation{%
  \institution{Swarthmore College}
  \streetaddress{500 College Ave.}
  \city{Swarthmore}
  \state{Pennsylvania}
  \country{USA}
  \postcode{19801-1390}
}

\author{Abdelrahman Abdelmonsef}
\authornotemark[1]
\email{ayahia1@swarthmore.edu}
\orcid{0000-0001-6952-2681}
\affiliation{%
  \institution{Swarthmore College}
  \streetaddress{500 College Ave.}
  \city{Swarthmore}
  \state{Pennsylvania}
  \country{USA}
  \postcode{19801-1390}
}

\renewcommand{\shortauthors}{Xiang and Abdelmonsef}

\begin{abstract}
    The differences in distributional patterns between benchmark data and real-world data have been one of the main challenges of using electroencephalogram (EEG) signals for eye-tracking (ET) classification. Therefore, increasing the robustness of machine learning models in predicting eye-tracking positions from EEG data is integral for both research and consumer use. Previously, we compared the performance of classifiers trained solely on finer-grain data to those trained solely on coarse-grain. Results indicated that despite the overall improvement in robustness, the performance of the fine-grain trained models decreased, compared to coarse-grain trained models, when the testing and training set contained the same distributional patterns \cite{vectorbased}. This paper aims to address this case by training models using datasets of mixed data complexity to determine the ideal distribution of fine- and coarse-grain data. We train machine learning models utilizing a mixed dataset composed of both fine- and coarse-grain data and then compare the accuracies to models trained using solely fine- or coarse-grain data. For our purposes, finer-grain data refers to data collected using more complex methods whereas coarser-grain data refers to data collected using more simple methods. We apply covariate distributional shifts to test for the susceptibility of each training set. Our results indicated that the optimal training dataset for EEG-ET classification is not composed of solely fine- or coarse-grain data, but rather a mix of the two, leaning towards finer-grain.
\end{abstract}



\maketitle

\section{Introduction}

Recently, machine-learning classifiers have found more and more consumer applications as well as "on the field" usage. For example, machine learning models can be used to identify shopping motives relatively early in the search process \cite{pfeiffer2020eye}, to detect workload strain in truck drivers \cite{lobo2016cognitive}, and to assess the diagnosis of many neurological diseases such as Autism Spectrum Disorder and Alzheimer's \cite{thapaliya2018evaluating,sotoodeh2021preserved,kang2020identification,deb2021trends}. Machine learning approaches have also shown adequate performance on computer vision bioinformatics applications, medical image analysis, Eye-Tracking (ET) analysis, and Electroencephalography (EEG) analysis \cite{qian2021two,qian2020multi,gu2020multi,xu2020multi,lotte2018review,qu2020multi,qu2020using,craik2019deep,qu2018eeg,roy2019deep,an2021mars,zhao2019bira,yao2022residual}.

As machine learning classification is used in more and more unfamiliar environments, it is increasingly important for these classifiers to be robust. In the EEG-ET research, recent work in this area has focused on determining what machine learning models are best equipped to predict eye position from EEG signals \cite{lotte2018review} as well as eliminating the noise associated with EEG data collection automatically \cite{plochl2012combining,oikonomou2020machine,roy2019machine,zhang2021eegdenoisenet,qu2022eeg4home,wang2022eeg}.

These approaches all examine robustness across a variety of underlying factors of data analysis. In this paper, We focus on the inherent distributional patterns and the underlying differences between finer-grain and coarser-grain data. For this paper, finer-grain data refers to data collected in a more complicated framework in an environment with more uncontrolled conditions. On the contrary, coarser-grain data refers to data collected in a more simplified collection format with more restrictions on the experiment's environment for the same task.

In medical research, fine-grain data collection methods have been explored with great success \cite{higginson2000ecological,plancher2012using}. That is the use of data from more complex, or finer-grained, data collection methods to test for simpler tasks. Oftentimes, systems developed in a coarse-grained "lab conditions" only work in controlled environments, causing difficulties when utilized in uncontrolled conditions \cite{wilson1993,marcotte2010neuropsychology}. Recently, fine-grain data approaches have shown successful results for Covid-19 contact tracing machine learning classifiers \cite{gomez2020simplistic}. In this paper, We want to emphasize the impact of data granularity as well as the beneficial effects of using fine-grain data in machine learning classification.

Previously, we showed that Machine Learning models trained on fine-grained data are more robust, meaning that they maintain more consistent accuracies, than models trained on coarse-grained data when tested on data of different complexities \cite{vectorbased}. We extend upon our previous findings by examining the idea of training on mixed fine- and coarse-grain datasets. To our knowledge, this approach was not tested with machine learning classifiers before. The goal of mixing datasets of varying data granularity is to improve upon the draw-back we previously found with training on purely fine-grain data. That is the model's performance when tested on data of similar complexity drops a significant amount when compared to purely coarse-grain data. We suspect that a mixed complexity training set may perform adequately across a broader range of testing distributional patterns, including different and similar complexities.

In this study, we train machine learning models for left-right eye-tracking classification using data from a mix of binary-classified (coarse-grain) and vector-based (fine-grain) collection frameworks. We then compare the results to models trained exclusively on either binary-classified or vector-based data. The goal is to expand upon our previous conclusions regarding the optimal method to increase both accuracy and robustness of machine learning classifiers using fine- and coarse-grain data. Robustness is determined by the accuracy after a covariate distributional shift. The distributional shift in combination with the different mixes of data complexity attempts to mimic realistic data which often contains varying distributional patterns. Since we had previously determined the superiority of fine-grain data in terms of robustness, we hypothesize that classifiers trained using more fine-grain data will attain higher accuracies after covariate distributional shifts are applied.

The purpose is to determine the "Goldilocks" or optimal training set composition of fine- and coarse-grain data for left-right gaze classification in terms of both accuracy and robustness as well as verify that fine-grain data performs better than coarse-grain data using a more encompassing experimental design. 

\begin{figure} [!t]
    \centering
    \includegraphics[scale = 0.7]{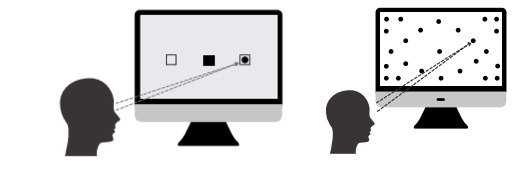}
    \caption{Schematic for the location of the cues on the screen in the PA (Left) and LG Tasks (Right) \cite{kastrati2021eegeyenet}}
    \label{fig:PALG}
\end{figure}

\section{Data}
We used the EEGEyeNet dataset due to its large size \cite{kastrati2021eegeyenet}. The dataset contains simultaneously-recorded EEG signals and Eye-Tracking data from three different experimental tasks. A summary table for the three tasks' metadata is in Appendix D. In this study, we used data from two experiments: pro-antisaccade (PA) and Large Grid (LG). In the PA trials, participants were asked to focus on a cue that appears on either the screen's center, horizontally left, or horizontally right, as shown in the left screen in Figure \ref{fig:PALG}. Gaze positions in PA were restricted to the horizontal axis and were binary-classified either left or right, relative to the screen's center. In the LG trials, participants were asked to fixate on dots presented one at a time at 26 different screen positions, as shown in the right screen in Figure \ref{fig:PALG}. Therefore, LG's framework enabled finer-grain data collection since it allowed more freedom for eye movements, and gaze positions were encoded in a two-dimensional format including both angle and amplitude, compared to the one-liner binary-encoded eye movements in PA. 

\begin{table*} [!htb]
    \centering
    \caption{The 20 different possible combinations of training and testing sets}
    \label{tab:ttcomps}
    \begin{tabular}{|c|cccccc|}
        \hline
        Training Set  & \multicolumn{6}{c|}{Testing Set}                                                                                                                                                 \\ \hline
        LG            & \multicolumn{1}{c|}{LG} & \multicolumn{1}{c|}{Mixed (20-80)} & \multicolumn{1}{c|}{Mixed (40-60)} & \multicolumn{1}{c|}{Mixed (60-40)} & \multicolumn{1}{c|}{Mixed (80-20)} & PA \\ \hline
        PA            & \multicolumn{1}{c|}{LG} & \multicolumn{1}{c|}{Mixed (20-80)} & \multicolumn{1}{c|}{Mixed (40-60)} & \multicolumn{1}{c|}{Mixed (60-40)} & \multicolumn{1}{c|}{Mixed (80-20)} & PA \\ \hline
        Mixed (20-80) & \multicolumn{3}{c|}{LG}                                                                           & \multicolumn{3}{c|}{PA}                                                      \\ \hline
        Mixed (40-60) & \multicolumn{3}{c|}{LG}                                                                           & \multicolumn{3}{c|}{PA}                                                      \\ \hline
        Mixed (60-40) & \multicolumn{3}{c|}{LG}                                                                           & \multicolumn{3}{c|}{PA}                                                      \\ \hline
        Mixed (80-20) & \multicolumn{3}{c|}{LG}                                                                           & \multicolumn{3}{c|}{PA}                                                      \\ \hline
    \end{tabular}
\end{table*}

\begin{figure} [t]
    \includegraphics[scale = 0.55]{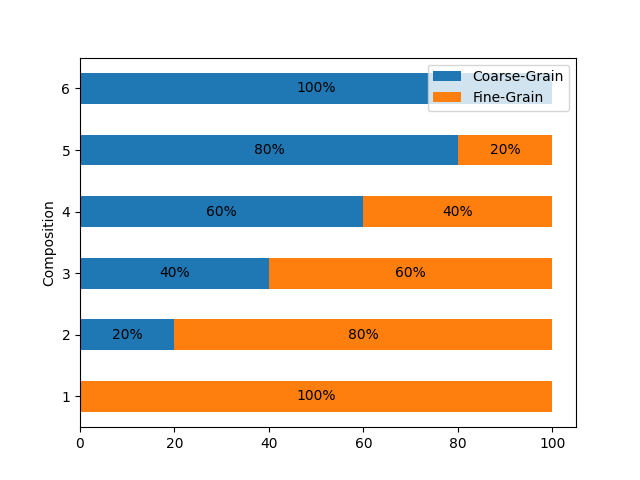}
    \caption{A description of the different compositions of fine- and coarse-grain data used in this study}
    \label{fig:comp}

\end{figure}

\section{Experiment Design}

The learning objective of the models trained in our experiment was to use EEG brain signals to predict the direction of a subject's gaze along the horizontal axis (whether they are looking to the left or the right). Although predictions for this task using the same dataset were previously made, they were performed exclusively using data from the PA tasks for training and testing  \cite{kastrati2021eegeyenet}.

In this paper, we train 13 classifiers using data that is composed of a mixture of both PA and LG data and perform a covariate distributional shift by comparing their performance based on their accuracy when tested on PA and LG data. The different compositions are described in Figure \ref{fig:comp}

\begin{figure}[b]
    \centering
    \includegraphics[scale = 0.4]{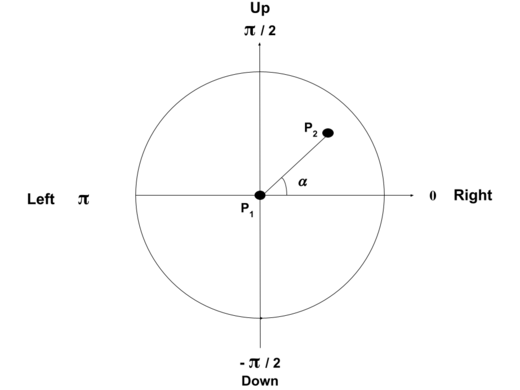}
    \caption{Illustration of angle $\alpha$. P1 represents the initial gazing position of the eye and P2 represents the end gazing position of the eye. The line between them represents the movement of the eye.}
    \label{fig:alpha}
\end{figure}

\subsection{Data Processing}
Given the classification nature of our learning problem, data from the LG task, encoded as Angle and Amplitude, should be transformed and relabeled into the expected format as left or right for training left-right classification models. The convention we used for the relabelling process was that for angle $\alpha$; when $|\alpha| < \frac{\pi}{2}$ the data is classified as right, otherwise the data is classified as left. This logic was confirmed by the dataset's authors in Appendix A and is shown in Figure \ref{fig:alpha}.

\subsection{Models Training}
As per EEGEyeNet authors' recommendation, we used the minimally preprocessed EEG data. Data processing was done using the NumPy library, and the model implementations were installed from the SKlearn library \cite{pedregosa2011scikit}.
From a broader perspective, models were trained and tested on 4 different combinations: models trained on mixed data and tested on PA data, models trained on mixed data and tested on LG data, models trained on PA data and tested on mixed data, and models trained on LG data and tested on mixed data. For each of the 4 combinations, the mixed data had 6 different compositions of PA and LG data, shown in Figure \ref{fig:comp}, which lead to 24 different combinations. However, 4 of these combinations are counted twice, leading to only 20 different combinations. A summary of the 20 unique combinations is shown in Table \ref{tab:ttcomps}.   


\section{Models}

\subsection{Machine Learning Models}
In this study, machine learning models operate on features extracted from the data rather than the data itself. Feature extraction has been applied in two steps. First, \cite{kastrati2021eegeyenet} applied a band-pass filter in frequencies in the range [8 - 13 HZ] on the acquired signals through all trials. This choice of frequencies is based on suggestions from \cite{foster2017alpha}. Following the filtering step, the Hilbert transform was applied, resulting in a complex time series from which targeted features were extracted for learning models. Since we are considering a classification problem, we experimented with classification-only models and models that can be applied to both classification and regression problems.     

\subsubsection{Linear Classifiers:} 
Linear classifiers gather discriminant classifiers that use linear decision boundaries between the feature vectors of each class \cite{lotte2018review}. In this paper, we will use Linear and Radial-Basis-function Support Vector Classifiers (SVC) and Normal and Shrinkage Linear Discriminant Analysis (LDA) classifiers. All four algorithms still perform well after several decades since they were first implemented in this field, with SVC outperforming other classifiers, especially for two-classes problems \cite{bashivan2016mental,lotte2015signal, lotte2018review}.

\subsubsection{Ensemble Classifiers} 

Ensemble (or voting) classifier is a machine learning classification algorithm that trains with different classification models and makes predictions through ensembling their predictions to make a stronger classification. These algorithms have been the gold standard for several EEG-based classification experiments \cite{qu2020multi}. In our study, we used Random Forest, XGBoost, GradientBoost, RUSBoost, and AdaBoost. 

\subsubsection{Naive Bayes and Decision Tree Classifiers}

Naive Bayes (NB) Classifier is the statistical Bayesian classifier \cite{duda1973pattern}. It assumes that all variables are mutually correlated and contribute to some degree towards classification. It is based on the Bayes' Theorem and is commonly used with high dimensional inputs. On the other hand, a decision tree is not a statistically based one; rather, it is a data mining induction technique that recursively partitions the dataset using a depth-first greedy algorithm until all data is classified with a particular class. Both NB and the decision tree are relatively fast and well suited for large data. Furthermore, they can deal with noisy data, which makes them well suited for EEG classification applications \cite{jadhav2016comparative}. 

\subsection{Deep Learning Models}
Deep learning is a subfield of machine learning algorithms in which computational models learn features from hierarchical representations of input data through successive non-linear transformations \cite{roy2019deep}. Deep learning methods, especially Convolutional Neural Network (CNN), performed well in several previous EEG band power (feature) based research \cite{qu2020multi}. Still, these methods have not demonstrated convincing and consistent improvements \cite{lotte2018review}. Given so and the expected high run time for such algorithms due to the dataset's large size, we only included a simple and relatively fast multi-layer perception neural network (MLP) in our experiment.

\begin{table*} [!h]

    \centering
    \caption{Models Trained on Mixed Data and Tested on Pro-Antisaccade Data}
    \label{tab:PATest}
    \begin{tabular}{|l|l|l|l|l|l|l|}
        \hline
        \multicolumn{1}{|c|}{Model} & \multicolumn{1}{c|}{PA} & \multicolumn{1}{c|}{Mixed (80-20)} & \multicolumn{1}{c|}{Mixed (60-40)} & \multicolumn{1}{c|}{Mixed (40-60)} & \multicolumn{1}{c|}{Mixed (20-80)} & \multicolumn{1}{c|}{LG} \\ \hline
        XGBoost                     & 96.7\%                  & 96.7\%                             & 95.7\%                             & 96.3\%                             & 95.5\%                             & 94.5\%                  \\ \hline
        GradientBoost               & 96.4\%                  & 96.3\%                             & 95.7\%                             & 95.7\%                             & 95.8\%                             & 94.2\%                  \\ \hline
        RandomForest                & 95.9\%                  & 95.7\%                             & 95.5\%                             & 95.1\%                             & 94.3\%                             & 93.3\%                  \\ \hline
        AdaBoost                    & 95.4\%                  & 94.8\%                             & 94.7\%                             & 94.2\%                             & 94.0\%                             & 93.1\%                  \\ \hline
        RUSBoost                    & 95.3\%                  & 95.2\%                             & 94.3\%                             & 93.7\%                             & 93.4\%                             & 93.4\%                  \\ \hline
        DecisionTree                & 94.4\%                  & 94.7\%                             & 94.1\%                             & 93.8\%                             & 93.4\%                             & 92.1\%                  \\ \hline
        MLP                         & 92.8\%                  & 93.9\%                             & 93.3\%                             & 92.4\%                             & 92.1\%                             & 91.3\%                  \\ \hline
        LinearSVC                   & 91.1\%                  & 90.4\%                             & 90.6\%                             & 90.5\%                             & 89.9\%                             & 89.9\%                  \\ \hline
        LDA                         & 90.6\%                  & 90.6\%                             & 90.2\%                             & 90.2\%                             & 90.0\%                             & 89.8\%                  \\ \hline
        sLDA                        & 90.3\%                  & 90.3\%                             & 90.1\%                             & 89.8\%                             & 90.1\%                             & 89.8\%                  \\ \hline
        KNN                         & 90.3\%                  & 89.3\%                             & 89.3\%                             & 88.1\%                             & 88.4\%                             & 87.9\%                  \\ \hline
        RBF SVC                     & 89.2\%                  & 89.2\%                             & 88.0\%                             & 87.3\%                             & 88.3\%                             & 88.1\%                  \\ \hline
        GaussianNB                  & 86.0\%                  & 85.4\%                             & 84.6\%                             & 83.9\%                             & 84.3\%                             & 82.8\%                  \\ \hline
        Average                     & 92.6\%                  & 92.5\%                             & 92.0\%                             & 91.6\%                             & 91.5\%                             & 90.8\%                  \\ \hline
    \end{tabular}
\end{table*}

\begin{table*} [!h]

    \centering
    \caption{Models Trained on Mixed Data and Tested on Large Grid Data}
    \label{tab:LGTest}
    \begin{tabular}{|l|l|l|l|l|l|l|}
        \hline
        \multicolumn{1}{|c|}{Model} & \multicolumn{1}{c|}{PA} & \multicolumn{1}{c|}{Mixed (80-20)} & \multicolumn{1}{c|}{Mixed (60-40)} & \multicolumn{1}{c|}{Mixed (40-60)} & \multicolumn{1}{c|}{Mixed (20-80)} & \multicolumn{1}{c|}{LG} \\ \hline
        XGBoost                     & 92.7\%                  & 94.1\%                             & 94.3\%                             & 95.4\%                             & 95.2\%                             & 95.4\%                  \\ \hline
        GradientBoost               & 91.8\%                  & 93.7\%                             & 93.4\%                             & 94.6\%                             & 94.9\%                             & 95.3\%                  \\ \hline
        RandomForest                & 90.9\%                  & 93.1\%                             & 93.1\%                             & 93.8\%                             & 94.0\%                             & 93.9\%                  \\ \hline
        AdaBoost                    & 89.2\%                  & 90.1\%                             & 91.5\%                             & 92.5\%                             & 93.0\%                             & 93.4\%                  \\ \hline
        MLP                         & 89.5\%                  & 91.2\%                             & 90.4\%                             & 91.2\%                             & 92.3\%                             & 92.9\%                  \\ \hline
        RUSBoost                    & 86.0\%                  & 90.4\%                             & 90.8\%                             & 92.5\%                             & 92.5\%                             & 92.7\%                  \\ \hline
        DecisionTree                & 88.0\%                  & 90.0\%                             & 90.1\%                             & 91.9\%                             & 91.6\%                             & 91.9\%                  \\ \hline
        sLDA                        & 90.3\%                  & 90.8\%                             & 91.5\%                             & 91.4\%                             & 90.9\%                             & 91.8\%                  \\ \hline
        LDA                         & 89.9\%                  & 90.6\%                             & 91.2\%                             & 90.9\%                             & 90.7\%                             & 91.6\%                  \\ \hline
        LinearSVC                   & 89.8\%                  & 91.2\%                             & 91.5\%                             & 91.5\%                             & 91.2\%                             & 91.1\%                  \\ \hline
        KNN                         & 89.3\%                  & 89.2\%                             & 88.8\%                             & 88.9\%                             & 89.9\%                             & 90.0\%                  \\ \hline
        RBF SVC                     & 84.8\%                  & 85.5\%                             & 85.8\%                             & 87.2\%                             & 87.4\%                             & 88.6\%                  \\ \hline
        GaussianNB                  & 83.4\%                  & 82.7\%                             & 82.8\%                             & 83.2\%                             & 87.1\%                             & 87.2\%                  \\ \hline
        Average                     & 88.9\%                  & 90.2\%                             & 90.4\%                             & 91.2\%                             & 91.6\%                             & 92.0\%                  \\ \hline
    \end{tabular}
\end{table*}

\begin{table*} [!h]

    \centering
    \caption{Models Trained on Pro-Antisaccade Data and Tested on Mixed}
    \label{tab:PATrain}
    \begin{tabular}{|l|l|l|l|l|l|l|}
        \hline
        \multicolumn{1}{|c|}{Model} & \multicolumn{1}{c|}{PA} & \multicolumn{1}{c|}{Mixed (80-20)} & \multicolumn{1}{c|}{Mixed (60-40)} & \multicolumn{1}{c|}{Mixed (40-60)} & \multicolumn{1}{c|}{Mixed (20-80)} & \multicolumn{1}{c|}{LG} \\ \hline
        XGBoost                     & 96.7\%                  & 95.4\%                             & 93.7\%                             & 93.3\%                             & 93.2\%                             & 92.7\%                  \\ \hline
        GradientBoost               & 96.4\%                  & 95.0\%                             & 93.2\%                             & 92.6\%                             & 92.7\%                             & 91.8\%                  \\ \hline
        RandomForest                & 95.9\%                  & 94.5\%                             & 92.2\%                             & 91.7\%                             & 91.7\%                             & 90.9\%                  \\ \hline
        AdaBoost                    & 95.4\%                  & 93.6\%                             & 91.0\%                             & 90.5\%                             & 90.1\%                             & 89.2\%                  \\ \hline
        RUSBoost                    & 95.3\%                  & 93.0\%                             & 89.5\%                             & 88.1\%                             & 87.2\%                             & 86.0\%                  \\ \hline
        DecisionTree                & 94.4\%                  & 92.2\%                             & 90.0\%                             & 88.6\%                             & 89.2\%                             & 88.0\%                  \\ \hline
        MLP                         & 92.8\%                  & 91.5\%                             & 89.5\%                             & 89.9\%                             & 90.4\%                             & 89.5\%                  \\ \hline
        LinearSVC                   & 91.1\%                  & 90.0\%                             & 88.7\%                             & 89.5\%                             & 90.4\%                             & 89.8\%                  \\ \hline
        LDA                         & 90.6\%                  & 89.9\%                             & 88.7\%                             & 89.2\%                             & 90.9\%                             & 89.9\%                  \\ \hline
        sLDA                        & 90.3\%                  & 89.6\%                             & 88.6\%                             & 89.2\%                             & 91.1\%                             & 90.3\%                  \\ \hline
        KNN                         & 90.3\%                  & 90.1\%                             & 89.6\%                             & 89.1\%                             & 90.1\%                             & 89.3\%                  \\ \hline
        RBF SVC                     & 89.2\%                  & 87.2\%                             & 85.5\%                             & 84.7\%                             & 86.0\%                             & 84.8\%                  \\ \hline
        GaussianNB                  & 86.0\%                  & 84.9\%                             & 84.0\%                             & 83.7\%                             & 84.9\%                             & 83.4\%                  \\ \hline
        Average                     & 92.6\%                  & 91.3\%                             & 89.6\%                             & 89.2\%                             & 89.8\%                             & 88.9\%                  \\ \hline
    \end{tabular}
\end{table*}

\begin{table*} [t]

    \centering
    \caption{Models Trained on Large Grid Data and Tested on Mixed}
    \label{tab:LGTrain}
    \begin{tabular}{|l|l|l|l|l|l|l|}
        \hline
        \multicolumn{1}{|c|}{Model} & \multicolumn{1}{c|}{PA} & \multicolumn{1}{c|}{Mixed (80-20)} & \multicolumn{1}{c|}{Mixed (60-40)} & \multicolumn{1}{c|}{Mixed (40-60)} & \multicolumn{1}{c|}{Mixed (20-80)} & \multicolumn{1}{c|}{LG} \\ \hline
        XGBoost                     & 94.5\%                  & 93.9\%                             & 94.1\%                             & 94.8\%                             & 95.9\%                             & 95.4\%                  \\ \hline
        GradientBoost               & 94.2\%                  & 93.7\%                             & 93.9\%                             & 94.5\%                             & 95.8\%                             & 95.3\%                  \\ \hline
        RandomForest                & 93.3\%                  & 93.0\%                             & 93.2\%                             & 93.1\%                             & 94.3\%                             & 93.9\%                  \\ \hline
        AdaBoost                    & 93.1\%                  & 93.0\%                             & 93.0\%                             & 92.9\%                             & 93.8\%                             & 93.4\%                  \\ \hline
        MLP                         & 91.3\%                  & 91.3\%                             & 92.0\%                             & 92.4\%                             & 93.6\%                             & 92.9\%                  \\ \hline
        RUSBoost                    & 93.4\%                  & 92.8\%                             & 92.5\%                             & 92.3\%                             & 93.2\%                             & 92.7\%                  \\ \hline
        sLDA                        & 89.8\%                  & 90.1\%                             & 90.1\%                             & 90.4\%                             & 92.6\%                             & 91.8\%                  \\ \hline
        DecisionTree                & 92.1\%                  & 91.9\%                             & 91.3\%                             & 91.1\%                             & 92.5\%                             & 91.9\%                  \\ \hline
        LDA                         & 89.8\%                  & 90.1\%                             & 89.9\%                             & 90.2\%                             & 92.3\%                             & 91.6\%                  \\ \hline
        LinearSVC                   & 89.9\%                  & 90.3\%                             & 89.9\%                             & 90.0\%                             & 92.0\%                             & 91.1\%                  \\ \hline
        KNN                         & 87.9\%                  & 87.8\%                             & 88.5\%                             & 88.8\%                             & 90.3\%                             & 90.0\%                  \\ \hline
        RBF SVC                     & 88.1\%                  & 87.4\%                             & 88.1\%                             & 88.5\%                             & 89.5\%                             & 88.6\%                  \\ \hline
        GaussianNB                  & 82.8\%                  & 83.4\%                             & 84.1\%                             & 84.9\%                             & 87.8\%                             & 87.2\%                  \\ \hline
        Average                     & 90.8\%                  & 90.7\%                             & 90.8\%                             & 91.1\%                             & 92.6\%                             & 92.0\%                  \\ \hline
    \end{tabular}
\end{table*}

\section{Results}

We trained and tested the machine learning models using datasets composed of both fine- and coarse-grain data (pro-antisaccade and "transformed" large grid) and compared the results. Thus, we had 20 combinations of training and testing datasets described in Table \ref{tab:ttcomps}.

Determining the Goldilocks composition of fine- and coarse-grain data for training is the main objective of this paper as it will indicate the optimal method of maintaining high accuracy and consistency. We also verified the results of fine- versus coarse-grain data by finding the accuracies of models when trained on PA/LG and tested on mixed data.

Results regarding the accuracies of select models as well as the average accuracy for each combination are provided in Tables \ref{tab:PATest}, \ref{tab:LGTest}, \ref{tab:PATrain}, and \ref{tab:LGTrain}. Table \ref{tab:MixTrain} is a summary of Tables \ref{tab:PATest} and \ref{tab:LGTest}.

\begin{table} [b]
    \centering
    \centering
    \caption{Average Accuracies of Models Trained on Mixed Data and Tested on Pro-Antisaccade/LG Data}
    \label{tab:MixTrain}
    \begin{tabular}{|l|l|l|l|}
        \hline
        \multicolumn{1}{|c|}{Data Composition} & \multicolumn{1}{c|}{PA} & \multicolumn{1}{c|}{LG} & \multicolumn{1}{c|}{Average}\\ \hline
        PA & 92.6\% & 88.9\% & 90.75\%\\ \hline
        Mixed (80-20) & 92.5\% & 90.2\% & 91.35\% \\ \hline
        Mixed (60-40) & 92.0\% & 90.4\% & 91.2\% \\ \hline
        Mixed (40-60) & 91.6\% & 91.2\% & 91.4\% \\ \hline
        \textbf{Mixed (20-80)} & 91.5\% & 91.6\% & 91.6\% \\ \hline
        LG & 90.8\% & 92.0\% & 91.4\% \\ \hline
    \end{tabular}
\end{table}

The average accuracy of models trained on PA and tested on mixed is 90.2\% as shown by the average of the averages in Table \ref{tab:PATrain}. The average accuracy of models trained on LG and tested on mixed is 91.3\% as shown by the average of the averages in Table \ref{tab:LGTrain}. The accuracy of models train on LG is also more compact and closer to the average accuracy. Clearly, finer-grain data is in general more applicable to real life data as it is more accurate across a broader range of distributional patterns. This confirms our previous findings across a more complete range of data complexity.

Note that although we did not include standard deviations in our tables, we did calculate them, but since they were insignificant (generally less than 0.1\%) we decided to not include them in our results. 

\section{Discussion}

Table \ref{tab:MixTrain} describes the average accuracy from 13 different machine learning model for left-right ET classification trained on data of different data complexity compositions. 

This paper utilizes the benchmark data from the EEGEyeNet dataset and 13 machine learning models to create left-right classifiers trained on data of varying compositions of fine- and coarse-grain data. The accuracies of the models are then tested using two testing sets with distinctly different distributional patterns (PA and LG). In this way, the effects of covariate distributional shifts are more apparent and provide insight on the optimal data complexity composition of fine- and coarse-grain data in order to reduce the impact of such phenomenon. The results also provide useful information on the types of machine learning classifiers that should be used for EEG-ET classification tasks.

\subsection{Finding the Goldilocks Composition}

Previously, we had concluded that machine learning classifiers trained purely on fine-grain data outperform machine learning classifiers trained purely on coarse-grain data in terms of robustness. This was confirmed by Tables \ref{tab:PATrain} and \ref{tab:LGTrain}. Therefore, we theorized that the ideal composition of data complexity would lean towards finer-grain data. 

The hypothesis is confirmed by Table \ref{tab:MixTrain}. The composition of 20\% coarse-grain data and 80\% fine-grain data produced the most consistently accurate results as shown by the average percentages. This suggests that the optimal training set for EEG-ET classification should not be created using data of solely one distributional complexity, but rather a mix of both fine- and coarse-grain data, leaning towards more fine-grain data. 

\subsection{Determining the Best Classifiers for EEG-ET Classification}
 
The consistently most accurate classifiers across Tables \ref{tab:PATrain}, \ref{tab:LGTrain}, \ref{tab:PATest}, and \ref{tab:LGTest} are XGBoost, GradientBoost, and RandomForest. Alternatively, the consistently worst performing classifiers were RBF SVC and GaussianNB. Therefore regardless of distributional complexity, XGBoost, GradientBoost, and RandomForest should be used to classify EEG-ET data.

\subsection{Future Recommendations and Improvements}
To further advance the work provided in this study, four steps are highlighted for future exploration. 

Although deep learning models were excluded due to inconsistent results \cite{lotte2018review}, the main reason was due to restraints in time and resources. As deep learning models, especially CNN and Attention \cite{vaswani2017attention,yi2022attention}, have shown promising results in regards to EEG-ET classification, it is important to thoroughly explore the effects of fine- versus coarse-grain data on the robustness of such models. 

Additionally, the angle value is currently the only indicator used to determine whether the saccade is towards the left or the right. The amplitude was completely ignored in data processing. The amplitude could be incorporated by using it as a weighting/scaling factor to indicate the left-right extension and solve this as a regression problem. Based on the predictions made by the regression value, we can then classify it as either left or right utilizing finer-grain data.

Furthermore, although we determined an approximately optimal training set, we are almost certain this is not the Goldilocks composition for mixed data complexity. Perhaps an application of gradient descent would be able to more precisely compute such a composition for each machine learning model and specific machine learning task. Additionally due to time restraints, we were unable to test our machine learning models trained on mixed distributional complexity against a broader range of testing sets like we did for exclusively fine- or exclusively coarse-grain trained classifiers in Tables \ref{tab:PATrain} and \ref{tab:LGTrain}.

Although our work discusses EEG-ET classification, the results can be potentially relevant to other machine learning models in other applications. We highly encourage further investigation on training machine learning models on data composed of different granularities across an array of other research topics.

\section{Conclusion}

The motivation behind this work was to extend upon our previous work on determining whether training machine learning models on finer-grain data leads to more robust models as well as to analyze the effects of training machine learning classifiers on datasets composed of mixed data complexity. We verified our previous findings across a broader range of distributional patterns as well as determined that ensemble methods are the most suitable machine learning classifiers for EEG-ET classification tasks. Furthermore we identified that EEG-ET machine learning classifiers seem to produce the most consistent results when the training set contains a mix of distributional patterns, leaning towards finer-grain data. We hope that future applications of practical EEG-ET interfaces utilize data of varying distributional patterns to increase classification accuracy and robustness. 

\section{Citations and Bibliographies}

\bibliographystyle{ACM-Reference-Format}
\bibliography{cite.bib}

\section{Appendices}

\subsection{Appendix A}

After emailing Ard Kastrati, Ard verified that for angle $\alpha$ in radians:

$\alpha = 0$ is right

$\alpha = \frac{\pi}{2}$ is down

$\alpha = \pi$ is left

$\alpha = -\frac{\pi}{2}$ is up
\\

\subsection{Appendix B}

The models were trained and tested on this environment settings:

OS: Mac 12.2.1

Cuda: 9.0, Cudnn: v7.03

Python: 3.9.0

cleverhans: 2.1.0

Keras: 2.2.4

tensorflow-gpu: 1.9.0

numpy: 1.22.1

keras: 2.2.4

scikit-learn  1.0.2

scipy 1.8.0

The total space occupied by the dataset on the device is 69.0574 GB, and the total time for training and testing was 30 mins on average. 

\subsection{Appendix C}

The code for data processing and evaluation is provided here: \url{https://github.com/ayahia1/KDD-ML}

\subsection{Appendix D}
The main reason for using EEGEyeNet in our study was its relatively large size. The table below presents the number of subjects, the number of sample data points, and the length of the recording time for each of the three experimental tasks in the EEGEyeNet dataset. The numbers below should prove the large size of this dataset.
\linebreak 
\linebreak
\textbf{Notes}: First, in the table, experimental tasks are referred to as "paradigms." Secondly, although the total number of subjects in the table is 486, some performed more than one experiment and, thus, referenced twice. The total number of unique subjects, when the dataset paper was first published, was 356. Finally, the recording time column contains the numbers rounded to the nearest hour. 

\begin{table} [!h]
    \centering
    \centering
    \caption{Metdata comparison of the 3 experimental paradigms}
    \label{tab:comparison}
    \begin{tabular}{|l|l|l|l|}
        \hline
        \multicolumn{1}{|c|}{Paradigm} & \multicolumn{1}{c|}{\# subjects} & \multicolumn{1}{c|}{\# samples} & \multicolumn{1}{c|}{Recording time}
        \\ \hline PA & 369 & 30842 & 38h
        \\ \hline Large Grid & 30 & 17830 & 8h 
        \\ \hline VSS & 87 & 31563 & 1h 
        \\ \hline Total & 486 & 80235 & 47h \\ \hline
    \end{tabular}
\end{table}

\end{document}